\documentclass[journal=langd5,manuscript=article]{achemso}

\usepackage[version=3]{mhchem} 

\author{Nupur Biswas}
\affiliation{Applied Material Science Division, Saha Institute of Nuclear Physics, 1/AF Bidhannagar, Kolkata 700064, INDIA}
\altaffiliation{Current address: Department of Physics, Indian Institute of Science, Bangalore 560012, INDIA}
\author{Alokmay Datta}
\affiliation{Surface Physics and Material Science Division, Saha Institute of Nuclear Physics, 1/AF Bidhannagar, Kolkata 700064, INDIA}
\email{alokmay.datta@saha.ac.in}

\title[An \textsf{achemso} demo]
  {`Coffee-ring' patterns of polymer droplets: chain entanglement effect}

\abbreviations{CL}

\begin{document}
\begin{abstract}
Dried droplets of polymer solutions of different molecular weights and concentrations leave various types of `coffee-ring' patterns. These patterns are consequence of contact line motion. We have observed for very low molecular weight the droplet periphery part contains spheroidal structures whereas central part contains continuous layer. For higher molecular weight contact line exhibits `stick-slip' motion. Whereas for very high molecular weight contact line delays enough to start motion and finally moves uninterruptedly leaving a continuous layer of polymer. We have explained this phenomena in terms of chain entanglement which is resultant of molecular weight and solution concentration. Depending on the entanglement chains can exhibit `granular' and `collective' behavior even when monomer numbers remain same within the droplets.
\end{abstract}

\section{Introduction}
When a drop of liquid solution dries the suspended particles deposit at the periphery leaving a thicker boundary. This phenomena is named as `coffee-ring' effect from the familiar stains of dried coffee drops. `Coffee-ring' stains are the outcome of self-assembly process of the solute molecules. The ring is formed at droplet periphery as evaporation being highest there solution carrying solutes move there to supply more solvent. Thus solute molecules accumulate and get deposited there to form `coffee-ring' \cite{Nature-389-827-1997}. Coffee-ring stains are very generalized phenomena \cite{AngewChem-51-1534-2012} and are observed for versatile materials, such as colloids \cite{PRE-61-475-2000}, polymers \cite{Langmuir-24-9102-2008}, biomolecules \cite{Langmuir-28-4039-2012,PRL-96-177801-2006}, liquid crystals \cite{PRE-60-638-1999}, nanoparticles \cite{JPCB-116-6536-2012,Langmuir-28-6163-2012}, carbon nanotubes \cite{Langmuir-26-2107-2010}. Although the mechanism of ring formation is driven by convection controlled by evaporation and surface tension of solvent \cite{Nature-389-827-1997}, there are other parameters such as nature of solute molecules \cite{Nature-476-308-2011,Langmuir-25-6934-2009}, the surface over which the droplet lies \cite{PRL-109-154501-2012,Langmuir-28-5331-2012,CPL-445-37-2007,Langmuir-24-9102-2008,Langmuir-20-3456-2004}, droplet size \cite{JPCB-114-5269-2010}, surrounding environment \cite{PRL-92-255509-2004,JACS-127-2816-2005} which also control this phenomena. Depending on these factors the droplets dry leaving the solute particles over the surface in various patterns. Also by adjusting Marangoni flow induced by surface tension gradient, coffee-ring formation can be completely suppressed and uniform deposition of solutes is obtained then \cite{JPCB-116-6536-2012,JPCB-110-7090-2006}.

Patterns formed by polymer molecules are our matter of interest for its scientific \cite{JCP-133-114905-2010} as well as technological relevance. Polymer droplets differ a lot from the usual coffee-ring stains which are formed from dispersions of colloidal particles. The drying mechanism and consequent self-assemble behaviour of polymer molecules are employed in different types of surface treatment and coating, films, spray drying for drug delivery \cite{PNAS-99-12001-2002}, inkjet printing \cite{NatMat-3-171-2004,Science-290-2123-2000}, data storage, surface-adhered proteins assays, DNA mapping technique \cite{Langmuir-21-3963-2005} and in soft patterning \cite{PhilTransRSocA-367-5157-2009}.

Polymers are long chain macromolecules whose dynamical behavior is governed by their entangled chains. Entanglement which is an unique property of polymers can be challenged by confining geometry. Within thin films of one dimensional confinement polymer molecules loss their inter-molecular entanglement and get layered \cite{Macro2}. Droplets are another example of confining space \cite{Science-332-1297-2011} where due to surface effects behaviour of polymer molecules deviates from their bulk phase characters \cite{CPL-539-157-2012,JColloidIntSc-318-225-2008}. Confinement may even lead to the crystallization of polymers \cite{PRL-92-255509-2004}.

Our aim was to investigate how molecular entanglement affects the self-organization behaviour of polymers within a macroscopic droplet. Here we report our results on molecular weight and concentration dependence of the coffee-ring stains left by the dried droplets of polymer.

\section{Experimental details}
We have worked with the droplets of atactic polystyrene (PS) of molecular weights 3700 (PS3), 18100 (PS18), 44000 (PS44), 114200 (PS1C), 560900 (PS5C) and 994000 (PS9C) gm/mole, all bought from Sigma-Aldrich, USA. PS solutions of varied concentrations were prepared by dissolving it in toluene (Merck, Germany) which is a `good' solvent \cite{Kawakatsu} for polystyrene \cite{ChemCommun-2275-1998}. Concentrations were maintained well below the critical `overlap concentration' \cite{Strobl} of each molecular weight. Millimeter sized droplets were deposited on Si (111) substrates (Ted Pella, Inc., USA) using micro liter syringe. Before depositing, Si (111) substrates were sonicated in toluene as it helps in the dewetting of polymer solutions \cite{EPJE-12-443-2003}.

All experiments were performed at ambient condition with temperature lying in the range 23$^\circ$C - 25$^\circ$C and relative humidity varied within a range of 35\% - 45\%. Optical microscope images were recorded using Olympus BX51 of Nanonics MultiView1000 microscope. Digital Image Analysis Software (DIAS) was used for recording both still and video images. ImageJ software (National Institute of Health, USA) was used for image analysis.

\section{Results and discussions}
Fig. 1 shows the optical microscope images of dried polystyrene droplets. We observe the patterns have changed with molecular weight and concentrations. These versatile patterns are the consequences of the contact line (CL) motion which is determined by the properties of solution and substrate. For droplets of all concentrations and molecular weight, at least within the range observed by us, the contact line moved towards the droplet center, although the details of motion varied significantly. This variation produced varieties of patterns. Here contact line motion exhibited a two step process \cite{Langmuir-21-3963-2005}. Initially the contact line remained pinned and after a certain time it started to recede. For low molecular weights (image A1 and A2 of Fig. 1) we observe a relatively faint boundary line and an almost uniform patch covering the whole area of droplet. However, for a higher range of molecular weights, multiple rings are observed (image B4, C4, D4 and E4 of Fig. 1). These multiple rings structure is absent at higher concentrations of higher molecular weights. Fig. 1-D6 and 1-E6 show the pattern from a dried drop of PS9C of concentrations 2mg/ml and 3mg/ml, respectively. These patterns have a very deep boundary line and some broad dark lines directed towards the droplet center.

In the drying process, rings are formed when contact line remains pinned. During the pinning period the solvent continues its evaporation leaving the solutes at the pinning point and deposited solutes form a ring. For low molecular weight (panel A of Fig. 1) the contact line starts its motion almost instantaneously after deposition of droplet and a faint boundary line is produced. Also the motion of contact line continues uninterruptedly without being pinned anywhere which leaves the solute polymer molecules over the whole area of the droplet. It produces a patch after complete evaporation which appears as a dewetting pattern at low concentration for diluteness of the initial concentration \cite{CPL-529-74-2012}. A closer look towards the droplet edge shows contact line leaves the solutes in the shape of small spheroidal structures (Fig. 2a). Polymer molecules of very low molecular weights remain in a coil shape within a dilute solution of good solvent. Their intra and inter molecular entanglements are also low. The molecules remain in a `granular' state. When droplet is formed from a solution containing such molecules within a highly evaporating solvent, the solvent starts rapid evaporation. The evaporation rate remains maximum near the droplet periphery due to the large area for curvature. As the solvent gets evaporated, the contact line moves inward rapidly leaving the polymer molecules on its way. At the same time, as the solvent evaporates, the concentration of the solution within droplet increases, the polymer chains go towards their `overlapping region', for which the central part of the droplet contains a continuous film of polymers even for a dilute solution of PS3 (Fig. 3a). Thus in the drying process, polymer molecules change their behaviour from granular to collective.

In case of droplets of higher molecular weights as entanglement is enhanced, the collective behaviour of polymer molecules dominates. Then contact line takes longer time to initiate its motion towards the center resulting a thick boundary line. The pattern shows ring-like structures (image B4, C4, D4 and E4 of Fig. 1). These rings appear when the CL stops during its `stick-slip' motion. This `stick-slip' motion is due to the sequential pinning-depinning of the CL and polymers get deposited whenever the CL remains pinned \cite{PRL-100-044503-2008}. We have also observed multiple branched rings (Fig. 1-A6) when whole part of the contact line does not move simultaneously \cite{Langmuir-18-3441-2002}.

The tendency of getting delayed to start motion of the contact line continues for more concentrated droplets of higher molecular weights (PS9C) which we have reported elsewhere \cite{AIP}. We have observed for concentrations of 2mg/ml onwards the whole part of the contact line moved simultaneously towards the center without being pinned anywhere and leaves a continuous film of polymer molecules over the whole area. Fig. 2 reveals the difference with the lower molecular weight droplets. Fig. 2 illustrates formation of continuous polymer film when molecular weight has enhanced maintaining same concentration i.e. monomer number remaining fixed. Hence this phenomena can be attributed towards the entangling behaviour of polymer chains which introduces a `collective' behaviour within the polymer molecules. The broad dark lines directed inward the droplet carries the signature of `collective' motion of strongly entangled polymers.


Fig. 4 illustrates the delay to start the contact line motion for two representative concentrations (2mg/ml and 3mg/ml) along with its exponential dependence
 \begin{equation}\label{eq-fit}
   t (M_w) = t^\prime - t^{\prime\prime}exp(-M_w/M_0)
\end{equation}
which leads to a saturation with increasing molecular weight ($M_w$). In Fig. 4 $t$ denotes time taken to initiate contact line motion and $t^\prime$ denotes saturation value for a given concentration. With increasing concentration the values of the parameters $t^\prime$, $t^{\prime\prime}$ and $M_0$ increase. As coffee-ring effect is a droplet-size dependent phenomena \cite{JPCB-114-5269-2010}, measurements were done for almost equal sized droplets.

\section{Conclusions}
We have observed self-assembling behaviour of polymer molecules within drying droplets. Depending on the molecular weight, hence entanglement, polymer molecules organize themselves. The patterns left after evaporation of solvent represent that organization. The pattern and drying mechanism reveals `granular' behaviour of polymer molecules at low molecular weight. Their collective behaviour emerges at high molecular weight. Thus patterns are governed by the entangling nature and confinement does not take over at this length scale.

These results show, besides the surface curvature, entanglement is the most influential factor in the evaporation process of polymer solutions within a droplet. It indicates molecular weight can be used a tuning parameter in drying process although more quantitative understanding of the physical process behind it is very much necessary to determine the shape of final pattern.

\begin{acknowledgement}

%

Author N. B. thanks Council of Scientific and Industrial Research (CSIR), Govt. of India and Director, SINP for awarding research fellowship.
\end{acknowledgement}

%
%


\newpage
\underline{FIGURE CAPTIONS}\\

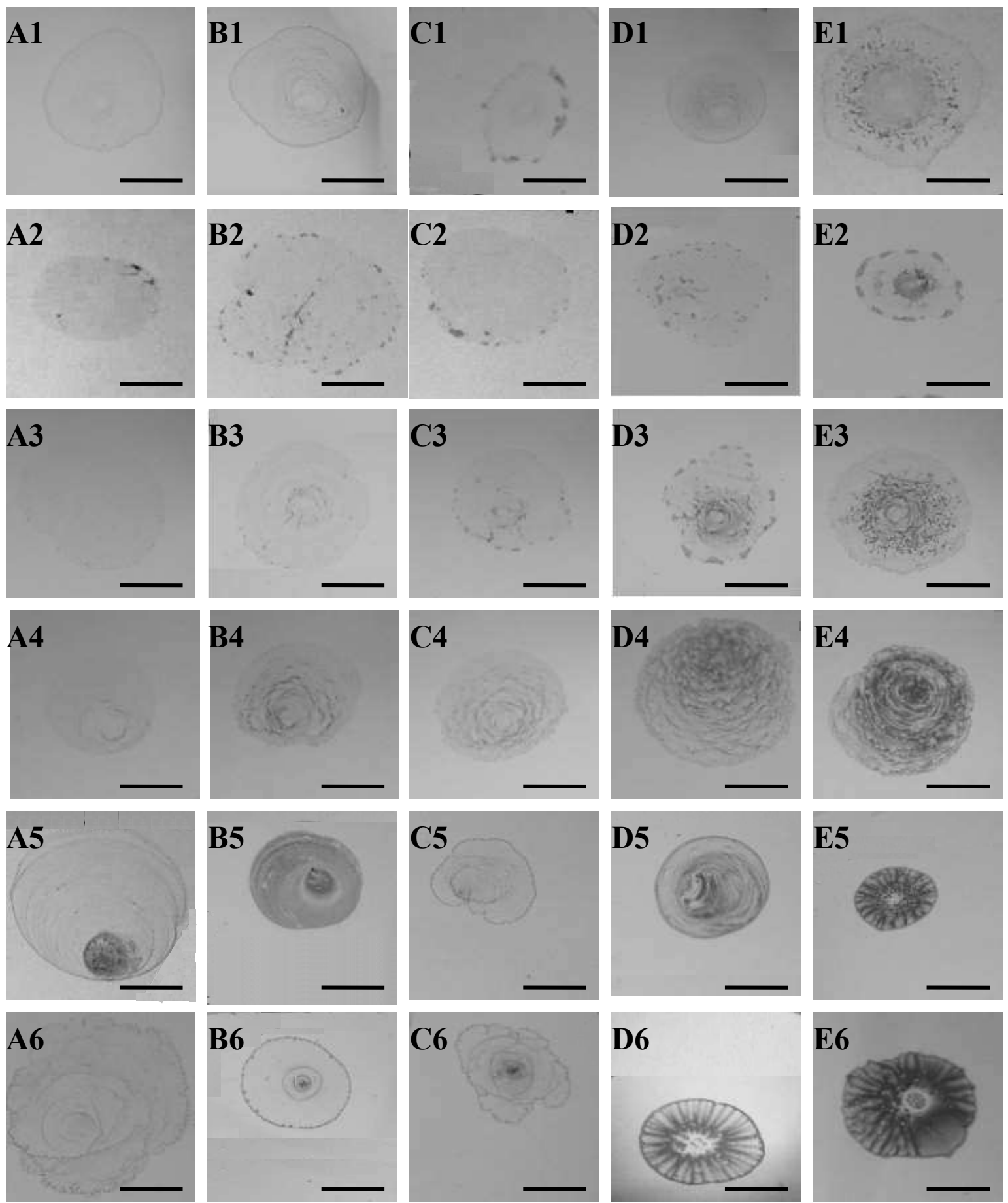
Figure 1: Optical microscope images of PS droplets of different molecular weights and concentrations. Sets A, B, C, D, E correspond concentrations 0.5mg/ml, 1mg/ml, 1.5mg/ml, 2mg/ml, 3mg/ml, respectively. Sets 1, 2, 3, 4, 5 correspond PS3, PS18, PS44, PS1C, PS5C, PS9C, respectively. All scale bars are 2 mm. \\

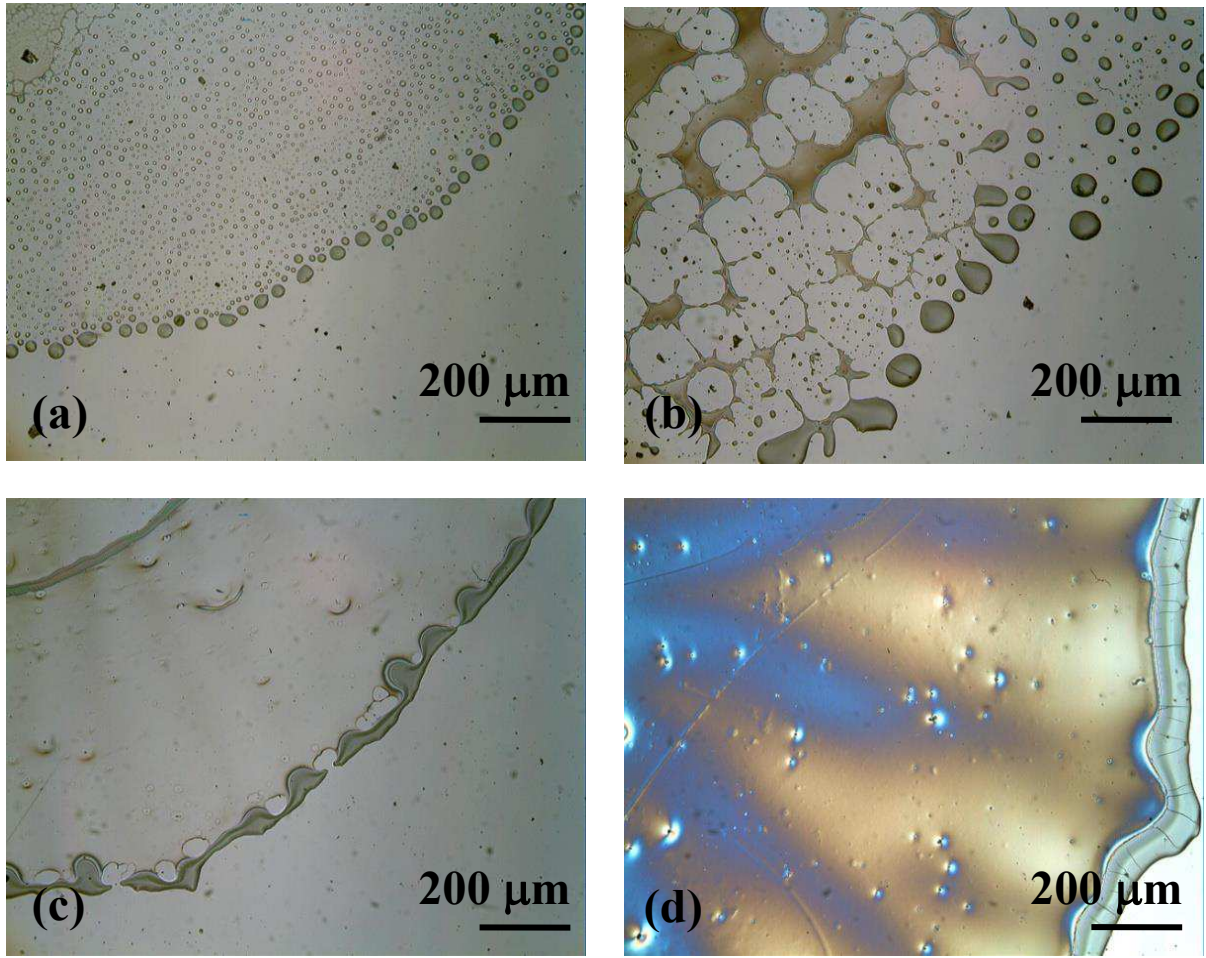
Figure 2: Magnified view of peripheral part of droplets: (a) PS3 of 1mg/ml, (b) PS3 of 3mg/ml, (c) PS9C of 1mg/ml and (d) PS9C of 3mg/ml. \\

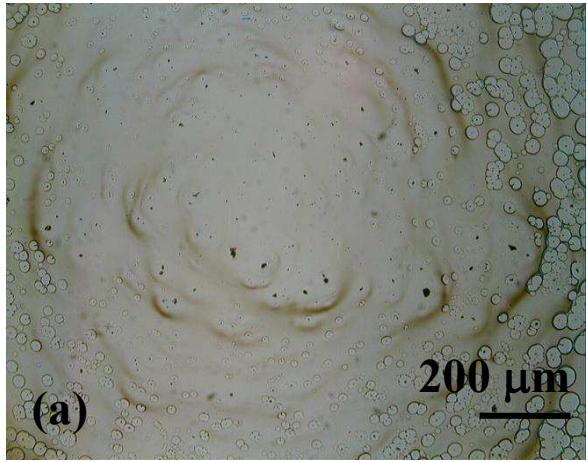
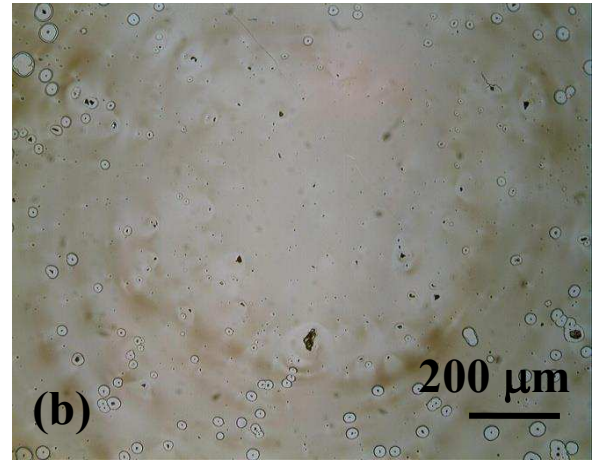
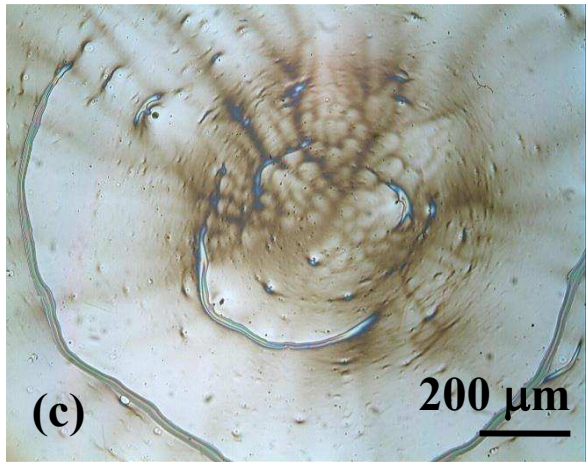
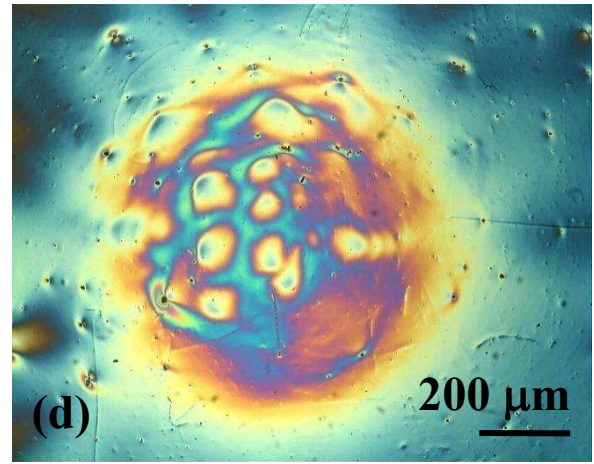
Figure 3: Magnified view of central part of droplets: (a) PS3 of 1mg/ml, (b) PS3 of 3mg/ml, (c) PS9C of 1mg/ml and (d) PS9C of 3mg/ml. \\

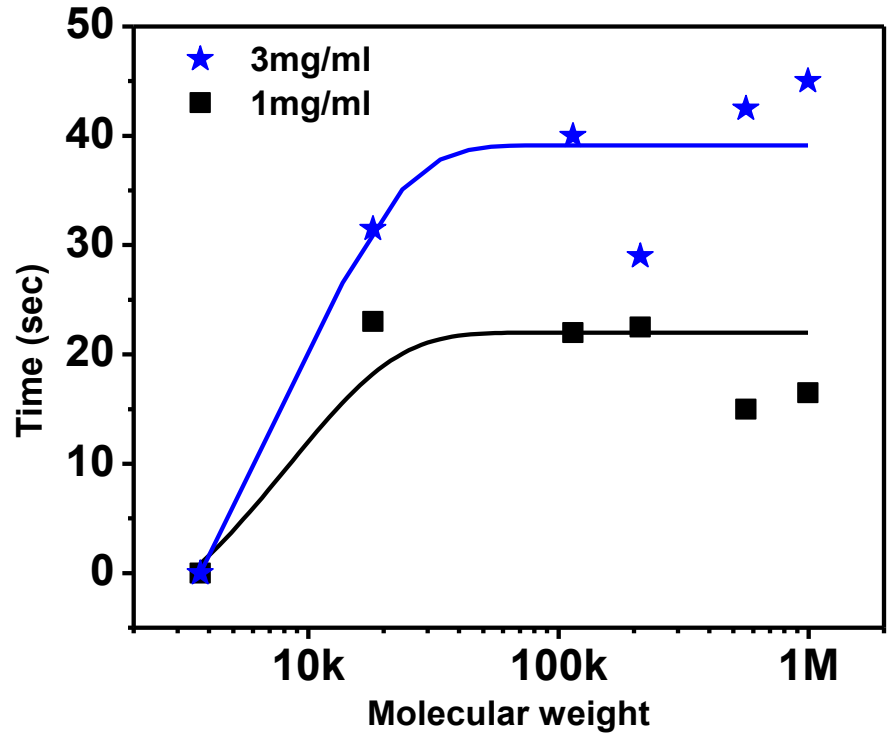
Figure 4: Time taken to initiate contact line motion with molecular weight. Scatter points are data, solid lines are fit by Eq. 1. \\

%
%

\end{document}